\documentclass[%
 aip,
 jmp,%
 amsmath,amssymb,
 reprint,%
]{revtex4-1}

\usepackage{graphicx}% Include figure files
\usepackage{dcolumn}% Align table columns on decimal point
\usepackage{bm}% bold math
\begin{document}

\preprint{AIP/123-QED}

\title{K and Mn co-doped BaCd$_{2}$As$_{2}$: a hexagonal structured bulk diluted magnetic semiconductor with large magnetoresistance}% Force line breaks with \\

\author{Xiaojun Yang}
 \affiliation{Department of Physics and State Key Laboratory of Silicon Materials, Zhejiang University, Hangzhou 310027, China}%Lines break
\author{Yuke Li}
 \affiliation{Department of Physics, Hangzhou Normal University, Hangzhou 310036, China}%Lines break
\author{Pan Zhang}
 \affiliation{Department of Physics and State Key Laboratory of Silicon Materials, Zhejiang University, Hangzhou 310027, China}
\author{Hao Jiang}
 \affiliation{Department of Physics and State Key Laboratory of Silicon Materials, Zhejiang University, Hangzhou 310027, China}
\author{Yongkang  Luo}
 \affiliation{Department of Physics and State Key Laboratory of Silicon Materials, Zhejiang University, Hangzhou 310027, China}
\author{Qian Chen}
 \affiliation{Department of Physics and State Key Laboratory of Silicon Materials, Zhejiang University, Hangzhou 310027, China}
\author{Chunmu Feng}
 \affiliation{Department of Physics and State Key Laboratory of Silicon Materials, Zhejiang University, Hangzhou 310027, China}
\author{Chao Cao}
 \affiliation{Department of Physics, Hangzhou Normal University, Hangzhou 310036, China}%Lines break
\author{Jianhui Dai}
 \affiliation{Department of Physics, Hangzhou Normal University, Hangzhou 310036, China}%Lines break
\author{Qian Tao}
 \affiliation{Department of Physics and State Key Laboratory of Silicon Materials, Zhejiang University, Hangzhou 310027, China}
\author{Guanghan Cao}
 \affiliation{Department of Physics and State Key Laboratory of Silicon Materials, Zhejiang University, Hangzhou 310027, China}
\author{Zhu-an Xu}
 \email{zhuan@zju.edu.cn}
 \affiliation{Department of Physics and State Key Laboratory of Silicon Materials, Zhejiang University, Hangzhou 310027, China}

\date{\today}% It is always \today, today,

\begin{abstract}
A bulk diluted magnetic semiconductor was found in the K and Mn
co-doped BaCd$_{2}$As$_{2}$ system. Different from recently
reported tetragonal ThCr$_{2}$Si$_{2}$-structured II-II-V
based(Ba,K)(Zn,Mn)$_{2}$As$_{2}$, the Ba$_{1-y}$K$_{y}$Cd$_{2-x}$Mn$_{x}$As$_{2}$
system has a hexagonal CaAl$_{2}$Si$_{2}$-type structure with the
Cd$_{2}$As$_{2}$ layer forming a honeycomb-like network. The Mn
concentration reaches up to its $x$ $\sim$ 0.4.
Magnetization measurements show that the samples undergo
ferromagnetic transitions with Curie temperature up to 16 K. With
low coercive field less than 10 Oe and large magnetoresistence
of about $-$70\%, the hexagonal structured
Ba$_{1-y}$K$_{y}$Cd$_{2-x}$Mn$_{x}$As$_{2}$ can be served as a promising
candidate for spin manipulations.
\end{abstract}

\pacs{75.50.Pp; 75.30.Kz; 85.75.-d; 75.30.Cr}

\maketitle

Diluted magnetic semiconductor (DMS) has long
been of great interest.\cite{Long JAP} The discovery
of ferromagnetism in Mn-doped GaAs was a milestone of the field of
DMS, which may not only exhibit substantial novel phenomena such as quantum Hall
effects, semiconductor lasers and single-electron charging, but also
bring about numerous applications in sensors, memories as well as
spintronics.\cite{1ohno Science,2I. Zutic RMP,3T. Dietl Science,4T.
Dietl Nat. Mat.,5 H. Ohno APL} However, traditional III-V based DMS
materials, represented by (Ga,Mn)As, are only available as thin
films prepared with nonequilibrium growth by
low-temperature molecular beam epitaxy (LT-MBE) technique, due to the
limited chemical solubility of manganese in GaAs ($<$ 1\%).
Accordingly, the sample quality sensitively depends on the
preparation methods and heat treatments,\cite{heat treatment}
which prevents accurate measurement and large-scale application of
DMS.\cite{13} High quality bulk DMS materials are therefore highly
desirable for both industrial applications and experimental research
purposes, for example, for the measurements of muon spin relaxation ($\mu$SR), nuclear magnetic
resonance (NMR) and neutron scattering.\cite{SR,LiZnAs15,BaZn2As2}
Moreover, the III-V based DMS materials lack of independently controlling of
local moment and carrier densities.

Recently, I-II-V based Li(Zn,Mn)As was theoretically proposed and
experimentally synthesized as a bulk DMS material with Curie
temperature (\emph{T}$_{C}$) of about 50 K.\cite{14,LiZnAs15} In this
system, charge and spin concentrations can be tuned separately by
controlling the contents of Li and Mn, respectively. Actually, only a limited
number of DMS systems own the ability to change the concentration of
acceptor and Mn impurities independently.\cite{PbSnMnTe,CdTeMn,ZnMnTe}
The chemical solubility of Mn is appreciably enhanced, and Mn
concentration reaches 15\% in bulk Li(Zn,Mn)As system. Inspired
by the rapid developement of iron-based
superconductors\cite{LaFeAsO,GdTh,Rotter122}, tetragonal
ThCr$_{2}$Si$_{2}$ structured  II-II-V
based (Ba,K)(Zn,Mn)$_{2}$As$_{2}$ was also reported to be a DMS
system with $T_{C}$ up to 180 K.\cite{BaZn2As2} Not only the bulk
DMS with high $T_C$, but also its structural proximity to
superconducting (Ba,K)Fe$_2$As$_2$ and antiferromangetic
BaMn$_2$As$_2$ may warrant the application for developing multi-layer
based functional devices.

BaCd$_{2}$As$_{2}$ possesses a hexagonal
CaAl$_{2}$Si$_{2}$-type structure (shown in Fig. 1(a,b)), which
belongs to P-3m1 (No.164) space group.\cite{BaCd2As2} The
Cd$_{2}$As$_{2}$ layers form a honeycomb-like network. The
honeycomb-like network has attracted more and more attention because
it is essential in recently extensively  investigated
topological insulators. Up to now, no study on the
physical properties of BaCd$_{2}$As$_{2}$ has been reported yet. In this paper,
we report our successful synthesis of a honeycomb-lattice bulk DMS
(Ba,K)(Cd,Mn)$_2$As$_2$, which shows ferromagnetic transition with $T_C$ up to 16 K. This material could be
the first bulk DMS system with a honeycomb-lattice. The system
features high manganese solubility up to $x\sim$0.4, which is not
only higher than traditional III-V DMS,\cite{1ohno Science} but also higher than the
newly discovered bulk DMS based on "111" LiZnAs, \cite{LiZnAs15} "122"
BaZn$_2$As$_2$, \cite{BaZn2As2} and "1111"LaZnAsO and LaCuSO systems.\cite{LaZnAsO, LaCuSO} Large negative magnetoresistance
($-$70\%) and small coercive field (less than 10 Oe) were observed
in this system. Its hexagonal structure and small coercive field
enable its applications in spintronics and other functional junction
devices in combination with topological insulators or
superconductors.

Polycrystalline samples
Ba$_{1-y}$K$_{y}$Cd$_{2-x}$Mn$_{x}$As$_{2}$ were synthesized by
solid state reaction method. All the starting materials, Ba rod, K
lumps, and the powders of Cd, Mn and As are of high purity ($\geq$
99.9\%). First, these materials were weighed according to the
stoichiometric ratio and put into crucibles, which were sealed in
evacuated quartz tubes, heated slowly to 1123 K, held for 10 hours
and then furnace-cooled to room temperature. After the first stage
of reaction, the products were ground, pelletized, put into
crucibles, sealed in evacuated quartz tubes and sintered at 1173 K
in vacuum for 33 hours, followed by cooling to room temperature
after switching off the furnace. Note that all the procedures
except for the tube sealing and heating were performed in a glove
box filled with high-purity argon.

Powder x-ray diffraction (XRD) was performed at room temperature using a PANalytical x-ray diffractometer (Model EMPYREAN) with a monochromatic CuK$_{\alpha1}$ radiation. The electrical resistivity was measured by four-terminal method. The temperature dependence of dc magnetization was measured on a Quantum Design Magnetic Property Measurement System (MPMS-5). The Hall coefficient was measured using a Quantum-Design physical property measuring system (PPMS). The thermopower was measured by using a steady-state technique.

\begin{figure}
\includegraphics[width=8cm]{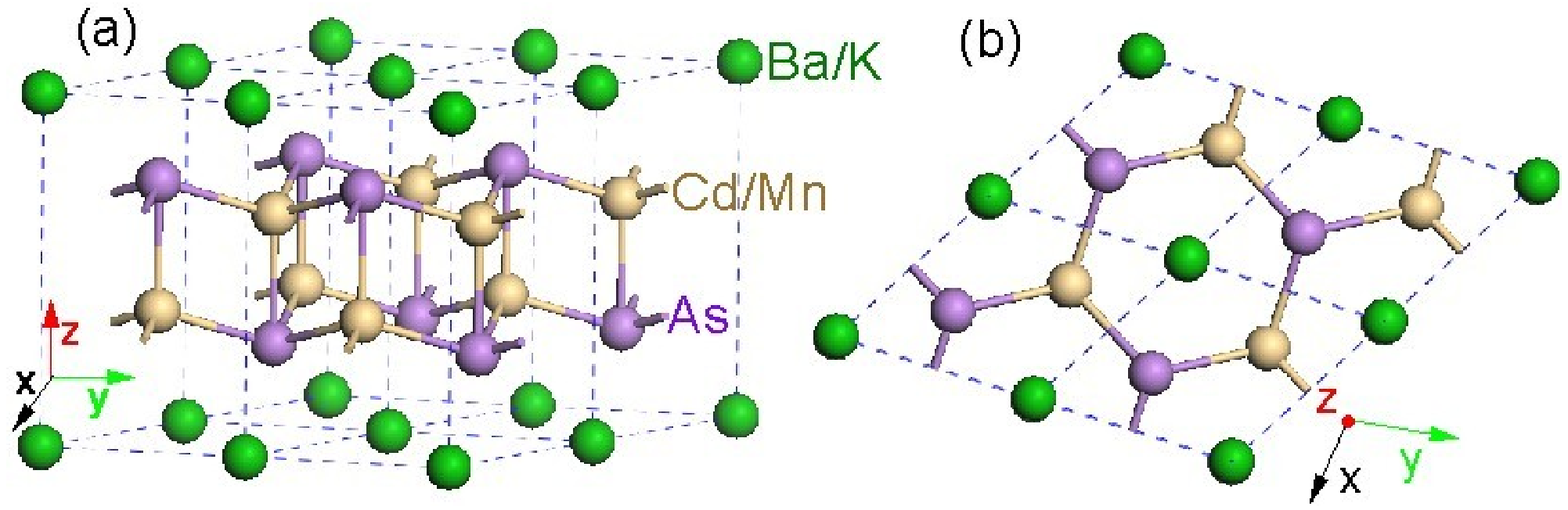}
\includegraphics[width=8cm]{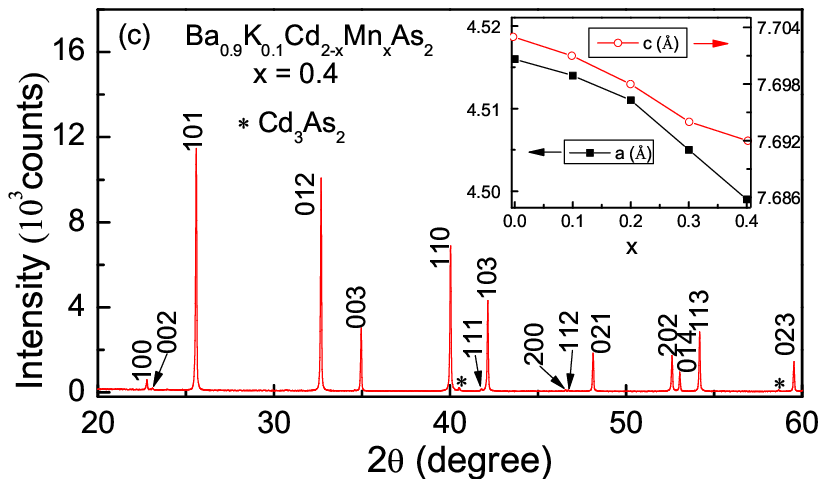}
\caption{\label{Fig.1} (color online)  (a,b), The crystal structure of (Ba,K)(Cd,Mn)$_{2}$As$_{2}$.
(c), A typical room-temperature
XRD pattern of Ba$_{0.9}$K$_{0.1}$Cd$_{2-x}$Mn$_{x}$As$_{2}$ (\emph{x} = 0.4)
sample indexed based on the P-3m1 (No.164) space group.  The minor peaks
marked by asterisks ($\ast$) are trace impurity phase of
Cd$_{3}$As$_{2}$ (0.7wt\% obtained by Rietveld refinement).
The inset of (c) exhibits the variation of lattice constants
$a$ (squares) and $c$ (circles) with Mn doping content (\emph{x}) in
Ba$_{0.9}$K$_{0.1}$Cd$_{2-x}$Mn$_{x}$As$_{2}$
system.}
\end{figure}

Fig. 1 shows a typical X-ray diffraction pattern of
Ba$_{0.9}$K$_{0.1}$Cd$_{2-x}$Mn$_{x}$As$_{2}$ for the \emph{x} =
0.4 specimen. The crystal structure is also sketched in Fig.1 (a)
and (b). All the peaks  can be well indexed based on the P-3m1
(No.164) space group with hexagonal CaAl$_{2}$Si$_{2}$-type
structure (the same as the undoped BaCd$_{2}$As$_{2}$ phase), except
for a few minor peaks assigned as impurity phase of
Cd$_{3}$As$_{2}$ (0.7wt\% obtained by Rietveld refinement). The
XRD patterns of other samples are basically the same, which indicates that
the samples are mainly composed of single phases. The
impurity phase Cd$_{3}$As$_{2}$ is a nonmagnetic
semiconductor due to the 4d$^{10}$ configuration of
Cd$^{2+}$,\cite{Cd2As3} thus the trace Cd$_{3}$As$_{2}$ impurity
phase will not affect the magnetic and transport properties of the
samples. Lattice parameters of the samples were obtained by
least-squares fit of more than 20 XRD peaks with the
correction of zero shift, using space group of P-3m1 (No.164). The
room temperature lattice constants $a$ and $c$ are 4.499
{\AA} and 7.692 \AA, respectively for the \emph{x} = 0.4 sample.
As shown in the inset of Fig. 1(c), both lattice constants $a$ and $c$ shrink
almost linearly with increasing Mn content in
Ba$_{0.9}$K$_{0.1}$Cd$_{2-x}$Mn$_{x}$As$_{2}$ (\emph{x} = 0, 0.1,
0.2, 0.3 and 0.4) system, indicating that the Mn ions were indeed
doped into the lattices, given that the ionic radius of the
Mn$^{2+}$ ion is known to be smaller than that of Cd$^{2+}$. The
Mn$^{2+}$ concentration of 20\% is remarkably high compared to the most III-V based ferromagnetic DMS systems and other
recently reported bulk ferromagnetic DMS systems.\cite{LiZnAs15,BaZn2As2} We
note that although bulk Zn$_{1-x}$Mn$_{x}$Te can be obtained with $x$ = 80\%, it is not ferromagnetic.\cite{Long JAP}

\begin{figure}
\includegraphics[width = 8cm]{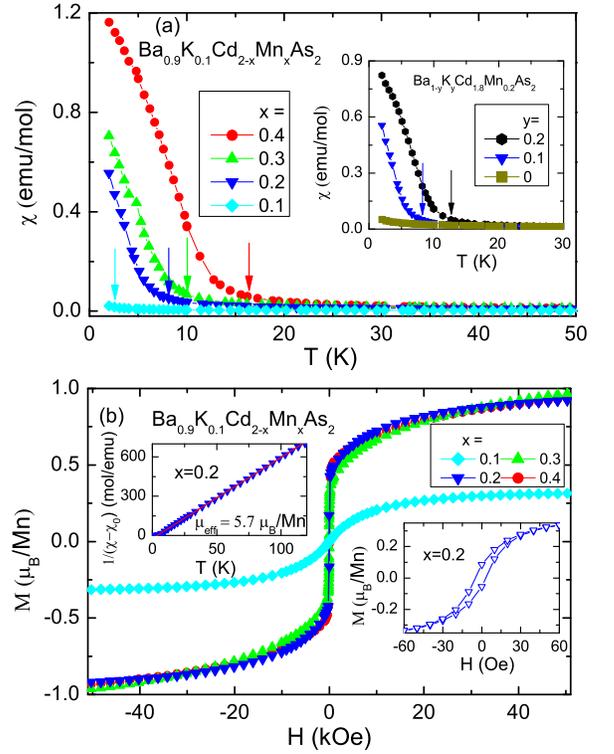}
\caption{\label{Fig.2}(color online)  (a),  Temperature dependence of dc magnetic
susceptibility measured under \emph{H} = 1 kOe for the
Ba$_{0.9}$K$_{0.1}$Cd$_{2-x}$Mn$_{x}$As$_{2}$ (\emph{x} = 0.1, 0.2, 0.3 and 0.4)
specimens.
The inset of (a): Temperature dependence of dc magnetic susceptibility measured
under \emph{H} = 1 kOe for the
Ba$_{1-y}$K$_{y}$Cd$_{1.8}$Mn$_{0.2}$As$_{2}$ (\emph{y} = 0, 0.1 and 0.2) samples.
(b),  Field dependence of magnetization measured at 2 K for
Ba$_{0.9}$K$_{0.1}$Cd$_{2-x}$Mn$_{x}$As$_{2}$ (\emph{x} = 0.1, 0.2, 0.3 and 0.4) samples.
The upper-left panel of (b) shows the 1/($\chi-\chi_{0}$) vs $T$ curve for $x$ = 0.2 sample.
Lower-right panel of (b) shows the  enlarged $M (H)$ curve for \emph{x} = 0.2 specimen.}
\end{figure}

\begin{figure*}
\includegraphics[width=14cm]{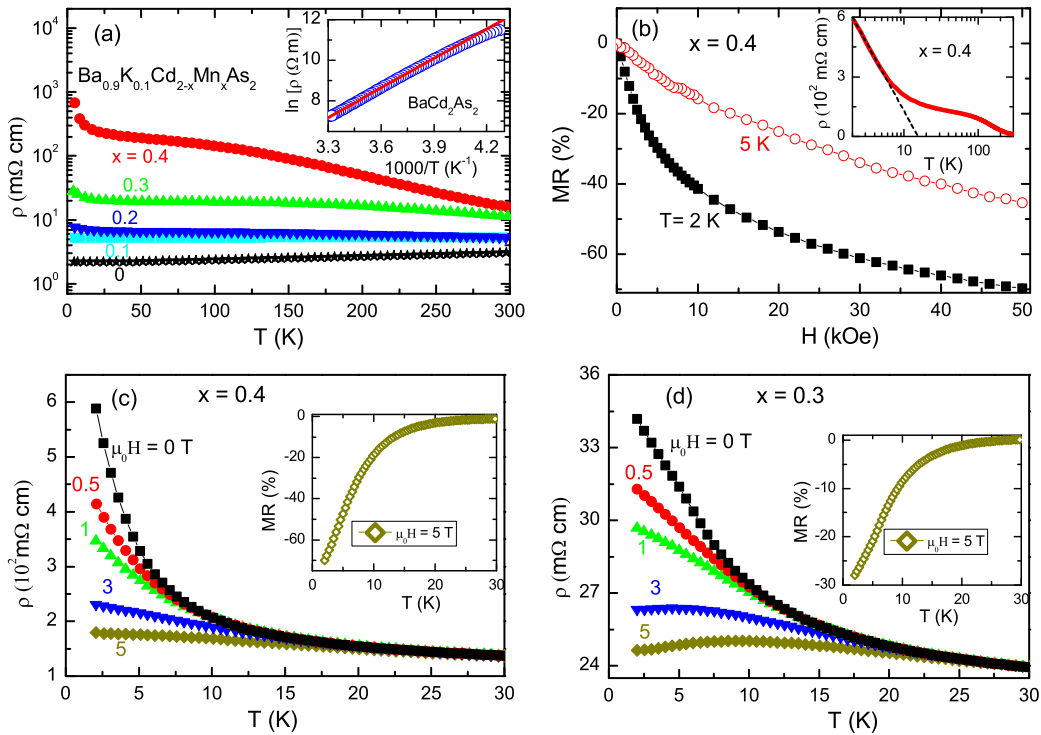}
\caption{\label{Fig.3}(color online)  (a), Temperature dependence
of resistivity of Ba$_{0.9}$K$_{0.1}$Cd$_{2-x}$Mn$_{x}$As$_{2}$
(\emph{x} = 0, 0.1, 0.2, 0.3 and 0.4) specimens. Inset displays
the resistivity of the parent compound BaCd$_{2}$As$_{2}$ in the
ln$\rho$ vs. $1/T$ plot. (b), Field dependence of
magnetoresistance measured at 2 K and 5 K for \emph{x} = 0.4
sample. Inset displays the resistivity of $x$ = 0.4 sample in the
log$T$ plot. (c,d): Temperature dependence of resistivity under
various external field of (c)\emph{x} = 0.4 and (d) \emph{x} = 0.3
samples. Inset of (c,d): Temperature dependence of MR under an
external field of $\mu_{0}$\emph{H} = 5 T for the samples with
(c)\emph{x} = 0.4 and (d) \emph{x} = 0.3.}
\end{figure*}

Figure 2(a) shows the temperature dependence of dc magnetic susceptibility of
Ba$_{0.9}$K$_{0.1}$Cd$_{2-x}$Mn$_{x}$As$_{2}$ under \emph{H} = 1 kOe
for samples with different Mn concentrations. No divergence can be
found between zero-field cooling (ZFC) and field-cooling (FC)
procedures. Apparently the samples for \emph{x} = 0.2, 0.3 and 0.4
become ferromagnetically ordered. The Curie temperatures ($T_{C}$,
denoted by arrows in Fig. 2(a)) are $\sim$ 8 K, 10 K and 16 K for
\emph{x} = 0.2, 0.3 and 0.4 samples, respectively.  For the sample
with \emph{x} = 0.1, an upward tendency of the susceptibility curve near
2 K is detected, suggesting possible ferromagnetic order below 2 K. The inset
of Fig. 2(a) exhibits the variation of temperature dependent dc
magnetic susceptibility of
Ba$_{1-y}$K$_{y}$Cd$_{1.8}$Mn$_{0.2}$As$_{2}$ under \emph{H} = 1
kOe for samples with different potassium concentrations. The \emph{T}$_{C}$
increases from 8 K to 13 K, as the potassium concentration \emph{y}
increases from 0.1 to 0.2. Since the major effects of Mn-doping and
K-doping are introduction of magnetic moment and charge carrier,
respectively; the above concentration dependence of $T_C$ is
consistent with the carrier-induced origin of the ferromagnetism
as describled by the Rudermann-Kittel-Kasuya-Yosida (RKKY) model
or its continuous-medium limit, that is, the Zener model.\cite{3T.
Dietl Science,p-ZnTe} As shown in the upper-left panel of Fig. 2(b),
the temperature dependence of magnetization
at temperature above $T_{C}$ can be well described by the modified
Curie-Weiss law, $\chi = \chi_{0} + C/(T-\theta)$, with the
effective paramagnetic moment values about 5.0-6.0 $\mu_{B}$/Mn,
as expected for fully magnetic individual Mn$^{2+}$ moments. The
samples with higher K and Mn doping levels were also made, but
their quality degrades substantially, and their $T_{C}$ could not be further
enhanced. The magnetic hysteresis loop $M(H)$ curves of
Ba$_{0.9}$K$_{0.1}$Cd$_{2-x}$Mn$_{x}$As$_{2}$ (\emph{x} = 0.1,
0.2, 0.3 and 0.4) specimens at \emph{T} = 2 K are shown in Fig.
2(b). The saturation moment reaches 0.46, 0.92, 0.97 and 0.94
$\mu_{B}$ per Mn atom at \emph{H} = 50 kOe, for the \emph{x} =
0.1, 0.2, 0.3 and 0.4 samples, respectively; which is comparable
with that in (Ga, Mn)As\cite{1ohno Science} and Li(Mn,
Zn)As\cite{LiZnAs15}. The ordered moment per Mn of the \emph{x} =
0.1 sample is only approximately half of that of other three samples,
suggesting that the moments may not become fully ordered in the
$x$ = 0.1 sample due to the low Mn concentration. The lower-right panel of
Fig. 2(b) exhibits a hysteresis loop for the \emph{x} = 0.2 sample
plotted for small field regions, showing a small coercive field of
less than 10 Oe. Such coercive field is even smaller than that of
Li(Mn,Zn)As (30-100 Oe),\cite{LiZnAs15} thus the material may be
more appealing in spin manipulations.

Resistivity measurements demonstrate that the parent compound
BaCd$_{2}$As$_{2}$ is a semiconductor. As shown in the inset of
Fig. 3(a), the thermal activation energy ($E_{a}$) obtained by
fitting with the thermal activation formula $\rho(T ) = \rho_{0}$
exp$(E_{a}/k_{B}T$) for the temperature range from 250 to 300 K is
about 0.42 eV. The temperature dependence of resistivity for the
Ba$_{0.9}$K$_{0.1}$Cd$_{2-x}$Mn$_{x}$As$_{2}$ (\emph{x} = 0, 0.1,
0.2, 0.3 and 0.4) specimens is shown in Fig. 3(a). The $x$ = 0
sample shows metallic conduction. For Mn-doped samples, clear
upturns in resisitivity are observed at low temperature, and become
more remarkable with an increasing of Mn doping level, which is
presumably due to magnetic scattering by Mn spins. This
kind of anomaly in resistivity can be observed in heavily-doped
semiconductors, which is similar with, for example, (Ga,
Mn)N\cite{GaN} and (Ba,K)(Zn,Mn)$_{2}$As$_{2}$ \cite{BaZn2As2}.
As shown in the inset of Fig. 3(b), the
low temperature resistivity for the sample with $x$ = 0.4
exhibits $-$log $T$ behavior. Usually, such a kind of $-$log $T$ behavior can be
attributed to Kondo effect\cite{CeNi2As2,CaNiGeH} or quantum correlations to the conductivity
in weakly localized regime.\cite{MIT_JPSJ,upturn-1_PRL,upturn-2_PE}
However, because the spin polarization of charge carriers will destroy
the Kondo effect, it may be caused by the weak localization in this system.
We then measured the magnetoresistance
(MR), defined as $[\rho(H)-\rho(0)]/\rho(0)$, for the $x$ = 0.4
sample by using isothermal measurements. As shwon in Fig. 3(b), MR
reaches up to $-$70\% at $T$ = 2 K and $\mu_0H$ = 5 T. The MR can
also be seen by measuring the temperature dependence of
resistivity under various magnetic fields (Fig. 3(c)). In the
absence of external magnetic field, the resistivity of the
\emph{x} = 0.4 specimen is 589 m$\Omega$ cm at 2 K, which
decreases rapidly with applied magnetic field, and reaches 179
m$\Omega$ cm under $\mu_{0}$\emph{H} = 5 T. The temperature
dependence of MR under an magnetic field of $\mu_{0}$\emph{H} = 5
T is depicted in the inset of Fig. 3(c), which reaches $-$70\% at
\emph{T} = 2 K and $\mu_{0}$\emph{H} = 5 T, consistent with the
isothermal measurement. An external magnetic field could reduce
the disorder among the local spins, and thus the magnetic
scattering will be suppressed, resulting in a negative
magnetoresistance. In Fig. 3(d), we display the temperature
dependence of resistivity under various magnetic fields for the
$x$ = 0.3 sample. Some interesting features can be observed.
First, when the applied field exceeds 3 T, a field
induced insulator to metal transition and a resistivity maximum
around $T_{C}$ can be observed. Second, under an applied external field,
resistivity starts to decrease even at the temperatures above
$T_{C} \sim$ 8 K, which could result from the suppression of
magnetic fluctuations near and above $T_{C}$. The inset of Fig.
3(d) shows the temperature dependence of MR for the sample with
\emph{x} = 0.3 under an magnetic field of 5 T, which reaches -28\%
at \emph{T} = 2 K. Large negative MR up to -22\% can also be found
for the sample with $x$ = 0.2 (not shown here), which indicates
that the large MR is a universal feature in this DMS system.
Materials exhibiting large magnetoresistance has been a focus of
interest because it can enlarge the sensitivity of read/write
heads of magnetic storage devices and thus maximize the
information density.\cite{CMR} The data indicate that
our system should be a magnetically doped semiconductor near
the metal-insulator transition regime.\cite{MIT_JPSJ}
In this regime, many interesting phenomena, such as field-induced
insulator-to-metal transition and large magnetoresistance, can be observed,
which should result from the complex interplay between quantum localization and carrier correlations.
Similar behaviors have been found in other reported systems in this regime. For example, the insulator-to-metal
transition, which is driven by the field-induced ordering of the Mn
spins, was found in p-type (Hg,Mn)Te.\cite{MIT_JPSJ,HgMnTe_PRL,HgMnTe_PRB} In (Ga,Mn)As system, a resistance
maximum near $T_{C}$ and negative magnetoresistance have been observed,
which results from the presence of randomly oriented ferromagnetic bubbles. \cite{GaMnAs Maximum and NMR,MIT_JPSJ}

\begin{figure}
\includegraphics[width = 8cm]{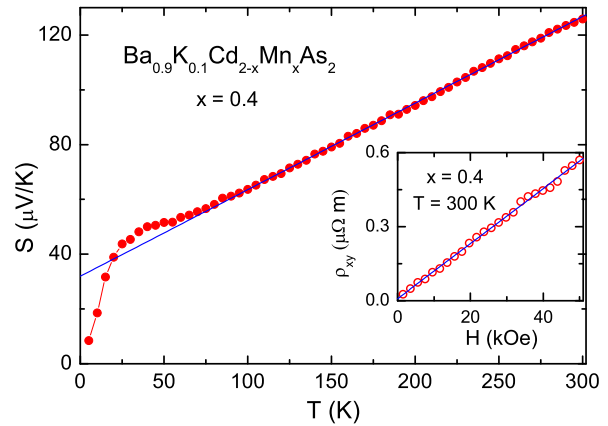}
\caption{\label{Fig.4}(color online) Temperature dependence of thermopower for the
Ba$_{0.9}$K$_{0.1}$Cd$_{2-x}$Mn$_{x}$As$_{2}$ (\emph{x} = 0.4) sample.
The inset displays field dependence of Hall resistivity of \emph{x} = 0.4
specimen at $T$ = 300 K}
\end{figure}

Figure 4 shows a typical temperature dependence of thermopower
($S(T)$) curve of Ba$_{0.9}$K$_{0.1}$Cd$_{2-x}$Mn$_{x}$As$_{2}$
for \emph{x} = 0.4 specimen. For the temperature range from 75 to
300 K, the $S(T)$ curve can be well fitted linearly, i.e. $S(T) =
CT$ with $C = 3.2\times10^{-7}$. In the free electron gas model,
the thermopower $S = \pi^{2}k_{B}^{2}T/3eE_{F}$, assuming that the
relaxation time is independent of free energy. Thus the fitting
derives the Fermi energy  $E_{F}$ = 0.077 eV. The $S(T)$ curve can
be well fitted by a free electron gas model, which is rather novel
considering that the \emph{x} = 0.4 sample exhibits
semiconductor behavior in the resistivity measurement. This kind of
phenomenon, which can also be found in other Mn doping level
samples (not shown here), should be further investigated. The
inset of Fig. 4 displays field dependence of Hall resistivity ($\rho_{xy}(H)$) of
$x$ = 0.4 specimen at $T$ = 300 K, which can also be fitted linearly. In a single-band
model, the Hall coefficient $R_{H}$ is associated with carrier
density ($n$) as $R_{H}$ = $1/ne$. We therefore estimate the
charge-carrier density to be 5.6 $\times$ 10$^{19}$/cm$^{3}$. Hall
coefficient is hard to measure at low temperature, because of
extremely large resistivity. The positive Hall coefficient and
thermopower demonstrate that hole type charge carrier is dominant
in the system, consistent with that 10\% K-for-Ba substitution
will introduces p-type charge carriers.

%\begin{figure}
%\includegraphics[width = 8cm]{Fig5.eps}
%\caption{\label{Fig.5} The calculated band structure of BaCd$_{2}$As$_{2}$ based on LDA.}
%\end{figure}

%To better understand the physical properties of BaCd$_2$As$_2$, we performed first-principles calculations. Fig. 5 (a) shows the band structure of BaCd$_2$As$_2$. The calculated band gap is a direct one of $\sim$ 140 meV located at $\Gamma$. The  magnitude of energy gap is approximately 1/3 of the experimental observation (0.42 eV), which is a well-known artifact of local density approximation and is expected to be fixed with GW correction. We have also calculated electron density of states (DOS) of the compound, and compared with the tetragonal BaZn$_2$As$_2$ in Fig. 5 (b).

%To better understand the structure and bonding, as well as the observed band gap properties
%of BaCd$_{2}$As$_{2}$, we performed first principle calculations. As shown in Fig.5,
%BaCd$_{2}$As$_{2}$ is a direct gap semiconductor with a separation
%between the top of the valence band and the bottom of the conduction band of about 0.16 eV.
%This magnitude of the energy gap is of the same order with the thermal activating gap (0.42 eV)
%derived experimentally.

In summary, we have successfully prepared a bulk hexagonal
structured ferromagnetic DMS system
Ba$_{1-y}$K$_{y}$Cd$_{2-x}$Mn$_{x}$As$_{2}$ by solid state
reaction. Via K-for-Ba substitution to supply hole-type carriers
and isovalent Mn-for-Cd substitution to introduce spins, the
density of local moment and charge carrier can be controlled
independently. The Mn concentration can reach up to
$\emph{x}$ $\sim$ 0.4, which is among the highest record of ferromagnetic DMS systems to our
knowledge.\cite{Long JAP,1ohno Science,LiZnAs15} Ferromagnetic order was
observed with \emph{T}$_{C}$ up to 16 K, and the saturation moment
of about 1 $\mu_{B}$/Mn. The \emph{T}$_{C}$ increases with
increasing both the Mn doping level (\emph{x}) and K concentration
(\emph{y}), which is roughly proportional to the spin and carrier
densities, respectively, consistent with the carrier-induced
origin of the ferromagnetism as describled by the RKKY model or
its continuous-medium limit, that is, the Zener model. With low
coercive field (less than 10 Oe) and high MR ($-$70\%), the system
provides an promising candidate for spin manipulation. In
particular, its hexagonal structure enables its combination with
topological insulators or superconductors to form devices with
novel functionalities.

We thank F.L. Ning for helpful discussions. This work is supported
by the National Basic Research Program of China (Grant Nos.
2011CBA00103, 2012CB927404 and 2012CB821404), NSF of China
(Contract Nos. 11174247 and 11190023), and the Fundamental
Research Funds for the Central Universities of China.

\end{document}